\documentclass[twocolumn]{aastex631}

\newcommand{\rev}[1]{\textcolor{black}{#1}}
\usepackage{comment}
\usepackage{array}
\usepackage{xcolor}
\usepackage{multirow}
\usepackage{xcolor}

\begin{document}

\title{Measurement of energy reduction by inertial Alfv\'en waves propagating through parallel gradients in the Alfv\'en speed}

\author[0000-0003-0124-8385]{Garima Joshi}
\affiliation{Columbia Astrophysics Laboratory, Columbia University, 550 West 120th Street, New York, NY 10027, USA}

\author[0000-0001-8093-9322]{Sayak Bose}
\affiliation{Princeton Plasma Physics Laboratory,  100 Stellarator Road, Princeton, NJ,  08540, USA}
\affiliation{Department of Astrophysical Sciences, Princeton University, NJ 08544, USA}

\author[0000-0002-5741-0495]{Troy Carter}
\affiliation{Department of Physics and Astronomy, University of California, Los Angeles, CA 90095, USA}

\author[0000-0002-1111-6610]{Daniel Wolf Savin}
\affiliation{Columbia Astrophysics Laboratory, Columbia University, 550 West 120th Street, New York, NY 10027, USA}

\author[0000-0002-6500-2272]{Shreekrishna Tripathi}
\affiliation{Department of Physics and Astronomy, University of California, Los Angeles, CA 90095, USA}

\author[0000-0002-6468-5710]{Stephen Vincena}
\affiliation{Department of Physics and Astronomy, University of California, Los Angeles, CA 90095, USA}

\author[0000-0001-7748-4179]{Michael Hahn}
\affiliation{Columbia Astrophysics Laboratory, Columbia University, 550 West 120th Street, New York, NY 10027, USA}

\begin{abstract}

We have studied the propagation of inertial Alfv\'en waves through parallel gradients in the Alfv\'en speed using the Large Plasma Device at the University of California, Los Angeles. The reflection and transmission of Alfv\'en waves through inhomogeneities in the background plasma is important for understanding wave propagation, turbulence, and heating in space, laboratory, and astrophysical plasmas. Here we \rev{present inertial Alfv\'en waves, under conditions relevant to solar flares and the solar corona. We find} that the transmission of the inertial  Alfv\'en waves is reduced as the sharpness of the gradient is increased. Any reflected waves were below the detection limit of our experiment and reflection cannot account for all of the energy not transmitted through the gradient. Our findings indicate that, for both kinetic and inertial Alfv\'en waves, the controlling parameter for the transmission of the waves through an Alfv\'en speed gradient is the ratio of the Alfv\'en wavelength along the gradient divided by the scale length of the gradient.  Furthermore,  our results suggest that an as-yet-unidentified damping process occurs in the gradient. 
\end{abstract}

%Unified Astronomy Thesaurus concepts :
%\keywords{Waves --- plasmas --- magnetohydrodynamic (MHD) --- solar corona --- Sun}

\section{Introduction} \label{sec:intro}

One of the most perplexing problems in modern solar physics is the mechanism responsible for heating the solar corona, the outer atmosphere of the Sun. The temperature of the corona is $\sim 200$ times greater than the underlying photosphere. Different physical theories exist to explain the phenomenon \citep{Cranmer:LRSP:2009}. One of the major theories posits that the heating is caused by Alfv\'en wave \citep{McIntosh:Nature:2011}. 

Strong damping of Alfv\'en waves has been observed at low heights in coronal holes, which are regions of the Sun with \rev{open magnetic field lines} extending into interplanetary space. The observed damping has been found to release sufficient energy to heat the solar corona \citep{Hahn:ApJ:2012, Bemporad:ApJ:2012,Hahn:ApJ:2013,Hara:ApJ:2019}. The Alfv\'en waves are generated by the jostling of flux tube footpoints \citep{Priest:ApJ:2000}. These Alfv\'en waves propagate outwards along the magnetic field lines (i.e., longitudinally). The waves are predicted to be partially reflected towards the Sun by the longitudinal gradient in the Alfv\'en speed along the magnetic field lines \citep{Moore:ApJ:1991,Matthaeus:ApJ:1999, Cranmer:ApJS:2005,Chandran:ApJ:2009}. The interaction of the reflected and outward propagating waves is thought to drive magnetohydrodynamic (MHD) turbulence, dissipating wave energy, and heating the solar corona \citep{Matthaeus:ApJ:1999, Cranmer:LRSP:2009}. Hence, understanding Alfv\'en wave propagation in the presence of an Alfv\'en-speed gradient is an important step in unraveling the heating mechanism of the solar corona. 
 
In our previous experimental work, we investigated the transmission and reflection of kinetic Alfv\'en waves for plasma parameters scaled to be representative of coronal holes \citep{Bose:ApJ:2019}. The kinetic Alfv\'en wave regime occurs when $v_{\rm A} < v_{T_{\rm e}}$, where $v_{\rm A}$ and $v_{T_{\rm e}}$ are the Alfv\'en speed and electron thermal speed, respectively. The Alfv\'en speed is given by $v_{\rm A} = B_{\rm 0}/\sqrt{\mu_{\rm 0} \rho}$. Here, $B_{\rm 0}$ is the ambient magnetic field; $\mu_{\rm 0}$ is the permeability of free space; $\rho =(n_{\rm i} m_{\rm i}+n_{\rm e} m_{\rm e})$ is the mass density of the plasma; $n_{\rm i}$ and $n_{\rm e}$ are the number densities of ions and electrons, respectively; and $m_{\rm i}$ and $m_{\rm e}$ are the ion and electron masses, respectively. The electron thermal speed is $v_{T_{\rm e}} = \sqrt{T_{\rm e}/m_{\rm e}} $, where $T_{\rm e}$ is the electron temperature in eV and $m_{\rm e}$ is the electron mass. Our kinetic regime experiments found that the wave energy transmitted through an Alfv\'en speed gradient was strongly reduced as the gradient became sharper compared to the wavelength. \rev{The wave energy that is not transmitted should be reflected or absorbed. Absorption could occur, for example, if the wave encountered a resonance in the plasma.} Our subsequent measurements of wave reflection in LAPD found that the reflection was surprisingly insufficient \citep{bose:ApJ:2024}. 

%In an ideal magneto-hydrodynamic(MHD) limit, the Alfv\'en waves carry energy and propagate mainly along the ambient magnetic field lines\citep{Alfv\'en:Nature:1942}. Alfv\'en waves are low-frequency magnetohydrodynamic waves, and Alfv\'en waves propagate below the ion cyclotron frequency, $\omega_{ci}=qB_0/m_i$. where q is the ion charge, and $m_i$ is the ion mass. 

Here, we extend our work to the inertial Alfv\'en wave regime, where $v_{\mathrm{A}} > v_{\mathrm{Te}}$. Our objective was to determine whether the transmission and reflection of inertial Alfv\'en waves differed from the previously measured kinetic Alfv\'en waves. In particular, since the Alfv\'en speed lies outside the electron energy distribution, we expected to greatly reduce the effects of Landau damping \citep{Lysak:JGR:1996, Theucks:PoP:2009}. 

In a solar physics context, inertial Alfv\'en waves are likely to be found in active regions where the strong magnetic field causes $v_{\mathrm{A}}$ to be large. For example, Alfv\'en waves have been hypothesized to accelerate electrons during solar flares \citep{Fletcher:aa:2008, McClements:aa:2009, Jess:LRSP:2023}. One mechanism by which they may accelerate electrons is through a turbulent cascade that occurs when downward propagating Alfv\'en waves reflect at the solar chromosphere. So, our measurements of the reflection and transmission of inertial Alfv\'en waves may have implications for understanding solar flares. 

The rest of the paper is organized as follows. In Section \ref{sec:exp}, we describe the experimental configuration. Section \ref{sec:result} describes the experiment results. This is followed by a discussion and summary in Section \ref{subsec:dis}.

\section{Experiment} \label{sec:exp}
\begin{figure*}[ht!]
\centering
\includegraphics[width=\textwidth]{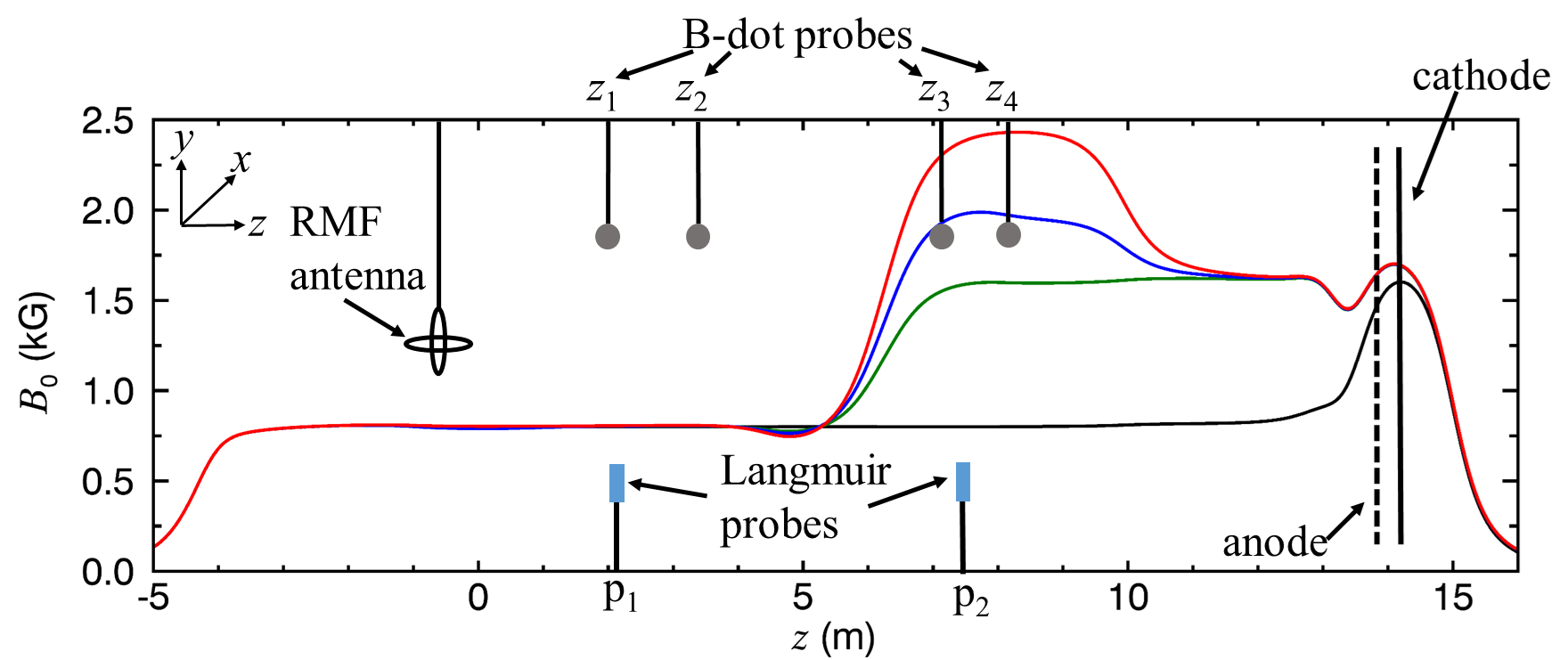}
\caption{Schematic of the experimental setup. The various magnetic field profiles used are shown in different colors. Black represents a uniform field of $B_{\rm 0}$ = 800 G. Not shown is the uniform field of 1600~G. Green, blue, and red represent 800~G on the low field side of our configuration and plateaus of 1600, 2000, and 2400~G on the high field side. The orthogonal loop antenna is on the low field side. Wave are launched toward the plasma source, i.e., the anode shown by the dashed line and the cathode shown by the solid line.  Also noted in the figure are the locations of the two Langmuir probes used to measure the plasma density and temperature and the four B-dot probes used to measure the wave magnetic field. 
\label{fig:schm}}
\end{figure*}

We conducted our experiment in the Large Plasma Device (LAPD) at the University of California, Los Angeles \citep{Gekelman:RSI:2016}. LAPD can generate a plasma column that is 18 m long and up to 60 cm in diameter. Our experiments were performed in a hydrogen plasma, for which the device was filled with neutral hydrogen at a pressure of $\sim 10^{-5}$ Torr.

The experimental schematic is shown in Fig. \ref{fig:schm}. A LaB$_6$ source \citep{Yuchen:rsi:2023} was used to generate the plasma by creating a discharge between the cathode and anode. Discharges (i.e., shots) were created every 3~s. Simultaneously with the onset of the discharge, hydrogen was puffed into the system near the plasma source for 19~ms. The discharge current ramped up to $\sim$ 5.5 kA in about $\sim 2$~ms, forming the main discharge. After 22~ms, the discharge voltage was turned off. \rev{When the discharge is turned off, the most energetic electrons rapidly leave the plasma, resulting in a cooler quiescent afterglow plasma, which was suitable for the inertial Alfv\'en waves studies reported here}. 

The plasma density, $n_{\rm e}$, and electron temperature, $T_{\rm e}$, were measured using Langmuir probes (p$_1$ and p$_2$ in Fig.~\ref{fig:schm}) and taking averages along an $x$ line through the plasma column at $y=0$. The density for p$_1$ was determined using the area of the probe tip. In the afterglow, the plasma density at p$_1$ was $n_{\rm e} = 1.6 \pm 0.3 \times 10^{11}$ cm$^{-3}$.  Here and throughout, unless otherwise state, all uncertainties are quoted at a $1\sigma$ statistical confidence level. The electron temperature $T_{\rm e}$ was found by sweeping the Langmuir probes from $-35$ to $+110$~V, from which we found $T_{\rm e} = 0.8 \pm 0.3$~eV. The density for p$_2$ was calibrated using a microwave interferometer. The plasma density at p$_2$ was $n_{\rm e} = 6.0 \pm 0.6 \times 10^{11}$ cm$^{-3}$. The electron temperature $T_{\rm e}$ was found by sweeping the Langmuir probes from $-35$ to $+50$~V, from which we found $T_{\rm e} = 1.0 \pm 0.3$~eV. The ion temperature is not routinely measured in LAPD, but previous studies have estimated it to be $T_{\rm i} \sim 0.5$~eV  \citep{Vincena:PoP:2013}.

%\vskip -12 pt
We used five axial magnetic field configurations which is along the $z$ direction. This included two uniform cases of $B_{\rm 0} = 800$ and 1600~G and three configurations with gradients rising from $B_{\rm 0}$ = 800~G before the gradient and increasing to 1600, 2000, and 2400~G after gradient, as shown by the colored curves in Fig.~\ref{fig:schm}. Throughout the paper, we will use the terminology before gradient, which refers to the region where the antenna is located, and after gradient, which refers to the region closer to the plasma source. 
% add something to define polarization

Alfv\'en waves were launched by an orthogonal loop antenna with frequencies that we varied from $(0.32-0.52)\omega_{\rm ci}$, where the ion cyclotron frequency is defined as $\omega_{\rm ci}=qB_{\rm 0}/m_{\rm i}$ and $q$ is the ion charge. Here and throughout, $\omega_{\rm ci}$ refers to the frequency for H$^+$. 

We launched left-hand circularly polarized shear Alfv\'en waves of 5 sinusoidal cycles using a rotating magnetic field (RMF) antenna \citep{Gigliotti:PoP:2009}. The RMF antenna consists of two rings, each with three turns of 0.25 cm diameter solid copper wire. The horizontal ring couples to the $y$-axis and is $\sim$ 8 cm diameter. The vertical ring couples to the $x$-axis and is $\sim$ 9 cm diameter. The rings were driven out of phase with each other by $\pi / 2$ . \rev{The antenna size determines the characteristic initial perpendicular wavelength of $\lambda_\perp \sim 9$ cm. However,  the wavelength grows to~$\sim 27$~cm as it propagates to the vicinity of the gradient at probe location $z_2$ (see Section~\ref{dispersion_kperp}), which we attribute to the changing plasma properties along the device.} The largest measured Alfv\'en wavelength along the magnetic field (i.e., parallel or longitudinal) was $\lambda_\| \approx$ 6 m, so that at least 3 wavelengths could be accommodated in a device. 

Starting at 4.3~ms after the onset of the discharge and continuing until 24.3~ms, every millisecond we launched a sequence of six wave trains. Each wave train was separated by $\sim 100$~$\mu$s and was at a different frequency. Here, we used six frequencies: $f=390$, 439, 488, 536, 585, and 634~kHz.

The Alfv\'en wave magnetic field $b$ was measured using B-dot probes at four axial locations. Each probe consists of three axes to measure the $b_{\rm x}$, $b_{\rm y}$, and $b_{\rm z}$ wave components. %The signal is then fed to the data acquisition system.  
The probes were mounted on linear translators that stepped through the LAPD cross section to obtain plane measurements. Wave field data were collected for planes at $z_1$ and $z_3$, covering 49 $\times$ 36 spatial locations. Line measurements along $x$, for $y=0$, were collected at $z_1$ = 223.65 cm, $z_2$ = 383.4 cm, $z_3$ = 743.85 cm, and $z_4$ =830.7 cm. The shot-to-shot variation in LAPD is very low, and the signal was averaged over 8 shots and digitized using a 16-bit data acquisition system. \rev{The probe remained stationary during data collection, then was moved to another position for the next measurement. This process was repeated throughout the plane.}
 
The motivation of our work is to understand the propagation of inertial Alfv\'en waves through a parallel $v_{\rm A}$-gradient. To that end, we adjusted the dimensionless plasma parameters in LAPD to match those in the Sun as closely as was experimentally feasible. Table~\ref{tab:my_label} compares the relevant parameters in LAPD to both coronal holes and and the low corona near the chromosphere of a solar flare. A detailed discussion comparing LAPD with coronal hole conditions has been given by \cite{Bose:ApJ:2019}. Here, we briefly review some specific parameters relevant to the present results. 

The wave amplitude $b$ relative to the background magnetic field $B_0$ was $\leq 10^{-5}$ in LAPD, but is about $b/B_{0} \approx 10^{-2}$ in coronal holes and $b/B_{0} \approx 10^{-1}$ in an active region during a solar flare. We chose to work with such a low value of $b/B_0$ to avoid nonlinear plasma effects. 

We have excited Alfv\'en waves with low frequency such that the parallel phase velocity of the wave is approximately equal to the Alfv\'en speed, $\omega/k_{\parallel} \approx v_{\mathrm{A}}$, as in the solar atmosphere. Here, $\omega$ is the angular frequency of the wave and $k_{\parallel}$ is the parallel wave number along the magnetic field lines. Our experiments, described below, verify that for $\omega/\omega_{\mathrm{ci}} \leq 0.5$, the Alfv\'en waves follow the dispersion relation for inertial waves. 

The perpendicular wave number $k_\perp$ in the solar corona has large uncertainties; but is assumed that ideal MHD holds and both  $k_\perp^2 \delta_{\rm e}^2$ and $k_\perp^2 \rho_{\rm i}^2$ are $\ll 1$, as is supported by recent observations at the base of the corona \citep{sharma:nature:2023}. This condition also holds in the the experiment. \rev{The value of $k_\perp$ is calculated using the Fourier-Bessel analysis discussed in Section \ref{dispersion_kperp} }. Here, $ \delta_{\rm e} =c/\omega_{\rm pe}$ is the collisionless electron skin depth, $c$ is the speed of light and $\omega_{\rm pe}=\sqrt{n_{\rm e} e^2/m_{\rm e}\epsilon_{\rm 0}}$ is the electron plasma frequency, where $e$ is the fundamental unit of electrical charge and $\epsilon_{\rm 0}$ is the permittivity of free space. The ion gyroradius is $\rho_{\rm i}=v_{\rm ti}/\omega_{\rm ci}$, where $v_{\rm ti}=\sqrt{T_{\rm i}/m_{\rm i}}$  is the ion thermal speed.

At the gradient, the Alfv\'en speed can change significantly. This is described using an inhomogeneity parameter, which is defined by $\lambda_\| /L_{\rm A}$, where $L_{\rm A}$ is the gradient scale length. \rev{The gradient scale length in the experiment and the coronal hole are similar.} Here, $L_{\rm A} = v_{\rm A}/v_{\rm A}'$ and $v_{\rm A}' = dv_{\rm A}/dz$ is the spatial derivative of $v_{\rm A}$ with \rev{respect} to $z$. The plasma medium is considered as uniform when $\lambda_\| /L_{\rm A} \ll 1$ and as non-uniform when $\lambda_\| /L_{\rm A} \geq 1$. 

The interplay between magnetic field pressure and thermal pressure is represented by the plasma beta parameter, $\beta=2\mu_{\rm 0} n_{\rm e}T_{\rm e}/B_{\rm 0}^2$. In the solar corona the magnetic pressure is generally much larger than the thermal pressure and $\beta \approx 10^{-3}$. The situation for solar flare is similar with \rev{ $\beta \lesssim 10^{-2}$}. In LAPD, this value varies from  ($10^{-4} - 10^{-5}$).

\begin{deluxetable}{lccc}[t]%\tabletypesize{\scriptsize}
\tablecaption{Dimensionless plasma parameters for inertial Alfv\'en waves in LAPD, coronal holes, and solar flares. \label{tab:my_label}}  
\tablehead{\colhead{Parameters} & \colhead{LAPD} & \colhead{Coronal holes} & \colhead{Solar flares}} 
\startdata 
${b}/B_{\rm 0}$                 & $10^{-5}$    & $10^{-2}$ & $10^{-1}$ \\ 
$k_\perp^2 \delta_{\rm e}^2$    & $\ll 1$            & $\ll 1$ & $\ll 1$ \\ 
$k_\perp^2 \rho_{\rm i}^2$      & $\ll 1$            & $\ll 1$ & $\ll 1$  \\ 
$\lambda /L_{\rm A}$            & 0 - 4              & $\geq 4.5$ & $\geq 20$ \\ 
$\omega/\omega_{\rm ci}$        & 0.15 - 0.52        &$\ll 10^{-5}$ & $10^{-9}$ \\ 
$\beta$                         & 10$^{-5}-10^{-4}$  &$\approx 10^{-3}$ & $ \lesssim 10^{-2}$ 
\enddata
\tablecomments{Flare parameters are for Alfv\'en waves at the corona/chromosphere boundary and based on the estimates of \citet{Fletcher:aa:2008}.}
\end{deluxetable}

\section{Results} \label{sec:result}

\subsection{Inertial Alfv\'en Wave}
\label{iaw}

\begin{figure*}[ht!]
\centering
\includegraphics[scale=0.34]{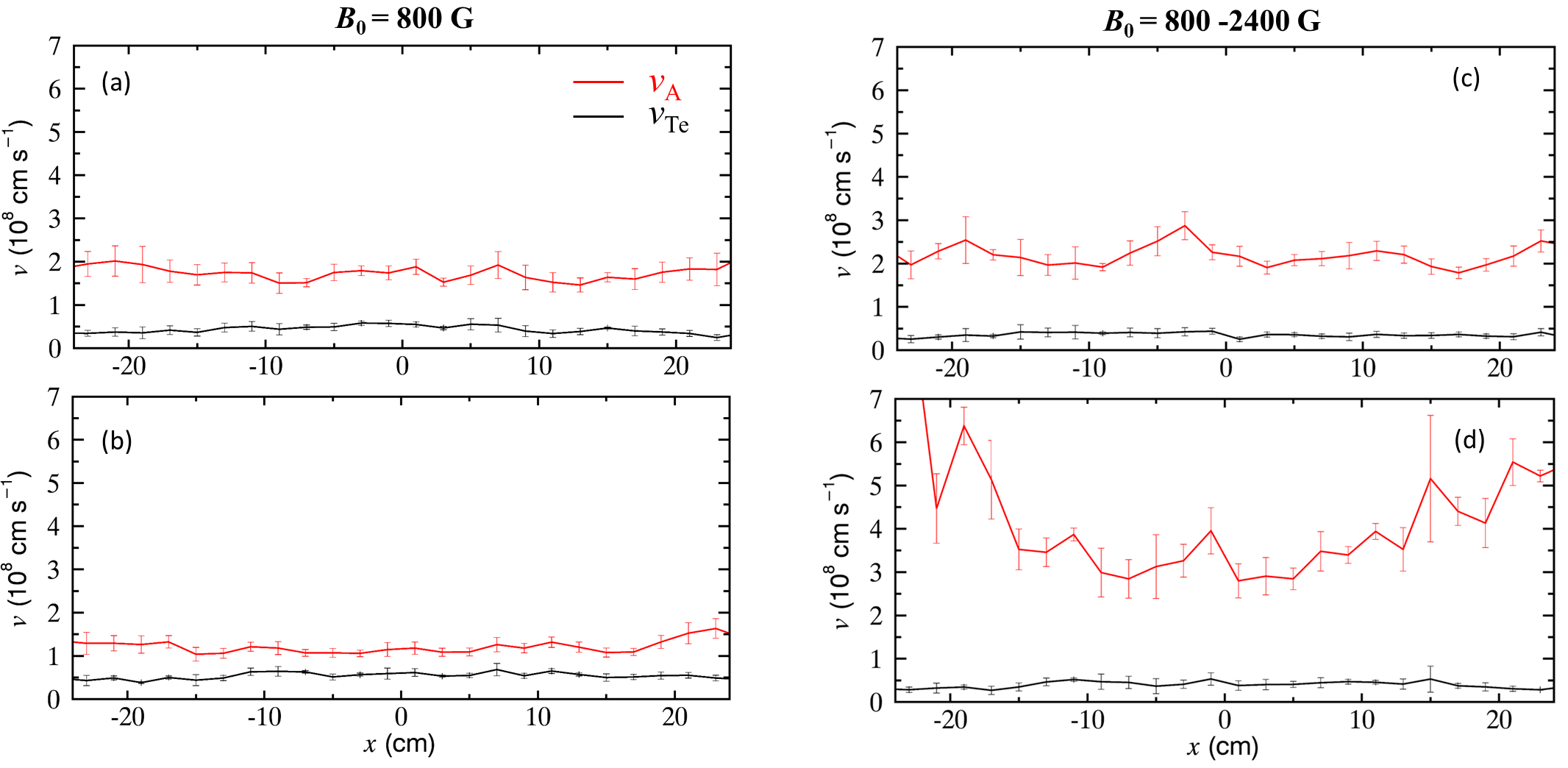}
\caption{Alfv\'en speed ($v_{\rm A}$ in red) and electron thermal speed ($v_{{T_{\rm e}}}$ in black) in  (a) and (b) for the flat field case at p$_1$ and p$_2$, respectively, and in (c) and (d) for the gradient case at p$_1$ and p$_2$, respectively.
\label{fig:alf_speed}}
\end{figure*}

In order to confirm that we are in the inertial Alfv\'en wave regime, we have calculated $v_{\rm A}$ and $v_{T{\rm e}}$. For this we have used  the $n_{\rm e}$ and $T_{\rm e}$ from the Langmuir probe measurements. These are shown in Fig.~\ref{fig:alf_speed} for the $B_{\rm 0} = 800$~G flat-field case and the $B_{\rm 0} = 800-2400$~G gradient case. Both $v_{\rm A}$ and $v_{T_{\rm e}}$ were determined by averaging along $x$ between $-10$ to $+10$~cm, which is where most of the wave energy is concentrated, as can be seen in Fig.~\ref{fig:wave_structure}. 
%The Langmuir probe locations P$_1$ and P$_2$ are referred to as  before and after the gradient, respectively, even for the flat-field case. 

For the flat-field case, before the gradient, $v_{\rm A}=1.7\pm 0.2\times10^8$~$\mathrm{cm\,s^{-1}}$ and $v_{\rm T_{\rm e}} =4.9 \pm 0.7\times10^7$~$\mathrm{cm\,s^{-1}}$, giving $v_{\rm T_{\rm e}}/v_{\rm A} =0.29$. After the gradient, $v_{\rm A}=1.2\pm 0.8\times10^8$~$\mathrm{cm\,s^{-1}}$ and $v_{\rm T_{\rm e}} =5.9 \pm 0.5\times10^7$~$\mathrm{cm\,s^{-1}}$, giving $v_{\rm T_{\rm e}}/v_{\rm A} =0.5$. For the $800-2400$~G gradient, before the gradient, $v_{\rm A}=2.2\pm 0.2\times10^8$~$\mathrm{cm\,s^{-1}}$ and $v_{\rm T_{\rm e}} =3.6 \pm 0.5\times10^7$~$\mathrm{cm\,s^{-1}}$, giving $v_{\rm T_{\rm e}}/v_{\rm A} =0.16$. After the gradient, $v_{\rm A}=3.2\pm 0.4\times10^8$~$\mathrm{cm\,s^{-1}}$ and $v_{\rm T_{\rm e}} =4.3 \pm 0.4\times10^7$~$\mathrm{cm\,s^{-1}}$, giving $v_{\rm T_{\rm e}}/v_{\rm A} =0.13$. These ratios demonstrate that our experiments satisfy the inertial Alfv\'en wave criterion of $v_{\rm A} > v_{\rm T_{\rm e}}$.

\subsection{Wave structure}
\label{subsec:wave_strct} 

\begin{figure*}[hbt!]
\centering
\includegraphics[scale=0.5]{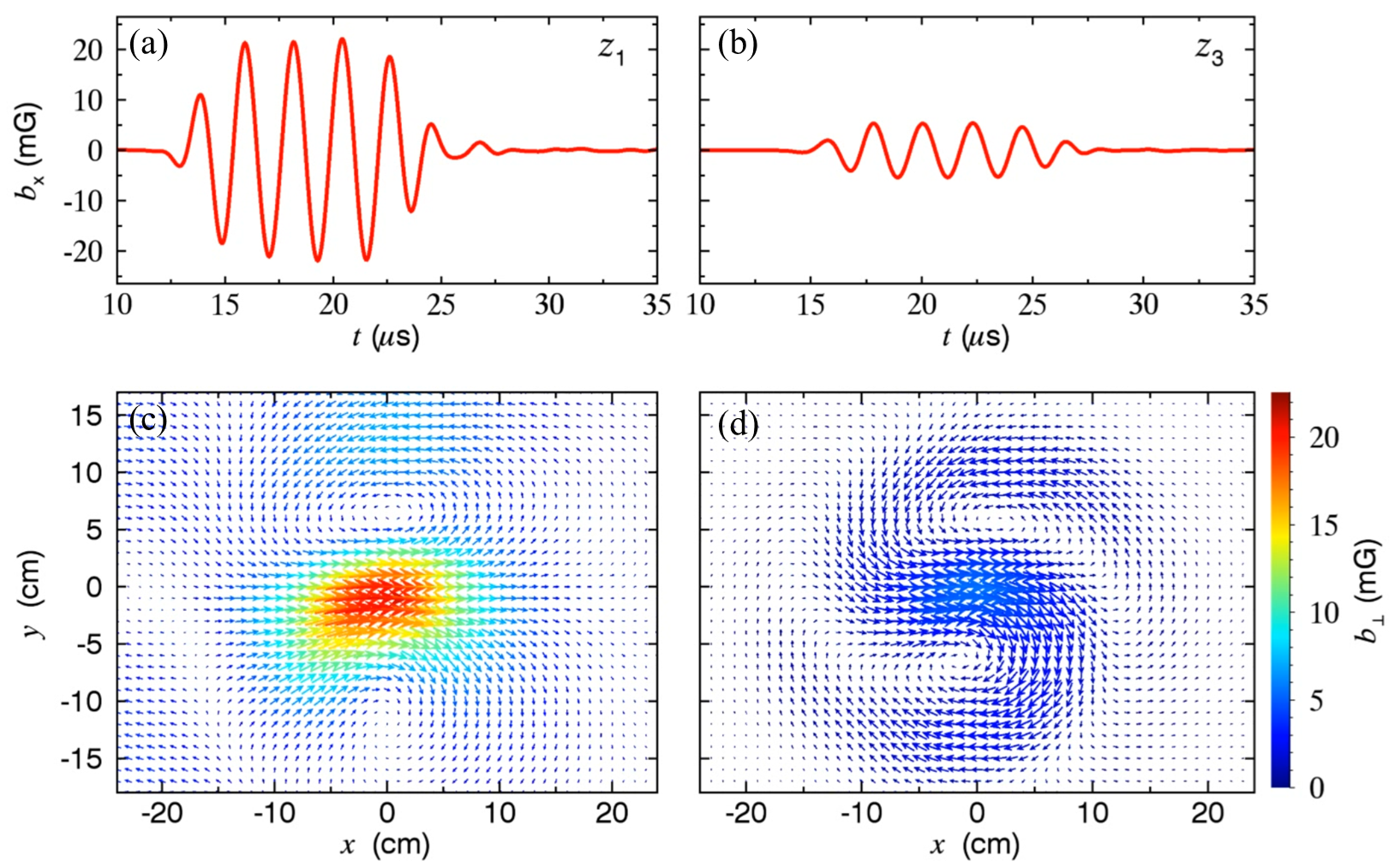}%width=\textwidth
\caption{Alfv\'en wave structure at two $z$ location for the $800-2000$~G gradient case for $x=y=0$. Panels (a) and (b) show the temporal profile of the $b_{x}$ component at $z_1$ and $z_3$, respectively. Panel (c) shows the wave spatial structure in an $xy$ cross section at $z_1$ at the time of the second peak in the temporal profile of the $b_{x}$ component in (a). Panel (d) shows the same but for $z_3$. See text for more details.
\label{fig:wave_structure}}
\end{figure*}

We have mapped the Alfv\'en wave structure in two $xy$ cross sections of LAPD: one before the gradient at $z_1$ and the other after the gradient at $z_3$. The wave temporal profile and spatial structure perpendicular to the ambient magnetic field are shown in Fig.~\ref{fig:wave_structure} for the $B_{\rm 0}=800-2000$~G gradient. The top two panels in the figure, (a) and (b), show the $b_x$ component of the wave magnetic field as a function of time at $x=y=0$ for a wave with $\omega = 0.36\omega_{\rm ci}$. Similar data were obtained in $xy$ cross sections of LAPD at $49 \times 35$ locations. The data in each cross-section plane were collected with a 1~cm step size. At each location, the data were averaged over eight shots. The bottom two panels in Figure~\ref{fig:wave_structure}, (c) and (d), show the perpendicular components of the wave magnetic field vector, $b_{\rm \perp} = \sqrt{b_{x}^2 + b_{y}^2}$. These two panels correspond to the time of the second peak in the time series data shown in panels (a) and (b). The direction of each arrow represents the orientation of the wave magnetic field at that location. The color of each arrow represents the magnitude of $b_{\rm \perp}$. The wave structure forms an $m=1$ mode, \rev{where $m$ is the azimuthal mode number.} The wave structure has two current channels $\sim 18$~cm apart at $z_1$ on the low field side and on the high field side $\sim 10$~cm apart at $z_3$.

\subsection{Alfv\'en wave dispersion relation}
\label{dispersion_kperp}
 
We launched Alfv\'en waves of frequencies $\omega$ from $(0.32-0.52) \omega_{\rm ci}$ and measured the wave signal along an $x$ line for $y=0$ at each of the four axial $z$ locations. Using a cross-correlation method, we calculated the time lag $\Delta t$ of the wave as it traveled from $z_1$ to $z_2$ before the gradient and from $z_3$ \rev{to}  $z_4$ and after the gradient. Since the distance between the probes is known, the phase velocity is given by $v_{\mathrm{ph},\|} = \Delta z / \Delta t$, which we then use to calculate $k_\| = \omega / v_{\rm ph,\|}$.  We have determined $v_{\rm ph,\|}$ for all magnetic field configurations considered here, both before and after the gradient. 

\begin{figure}[t!]
\centering
\includegraphics[scale=0.36]{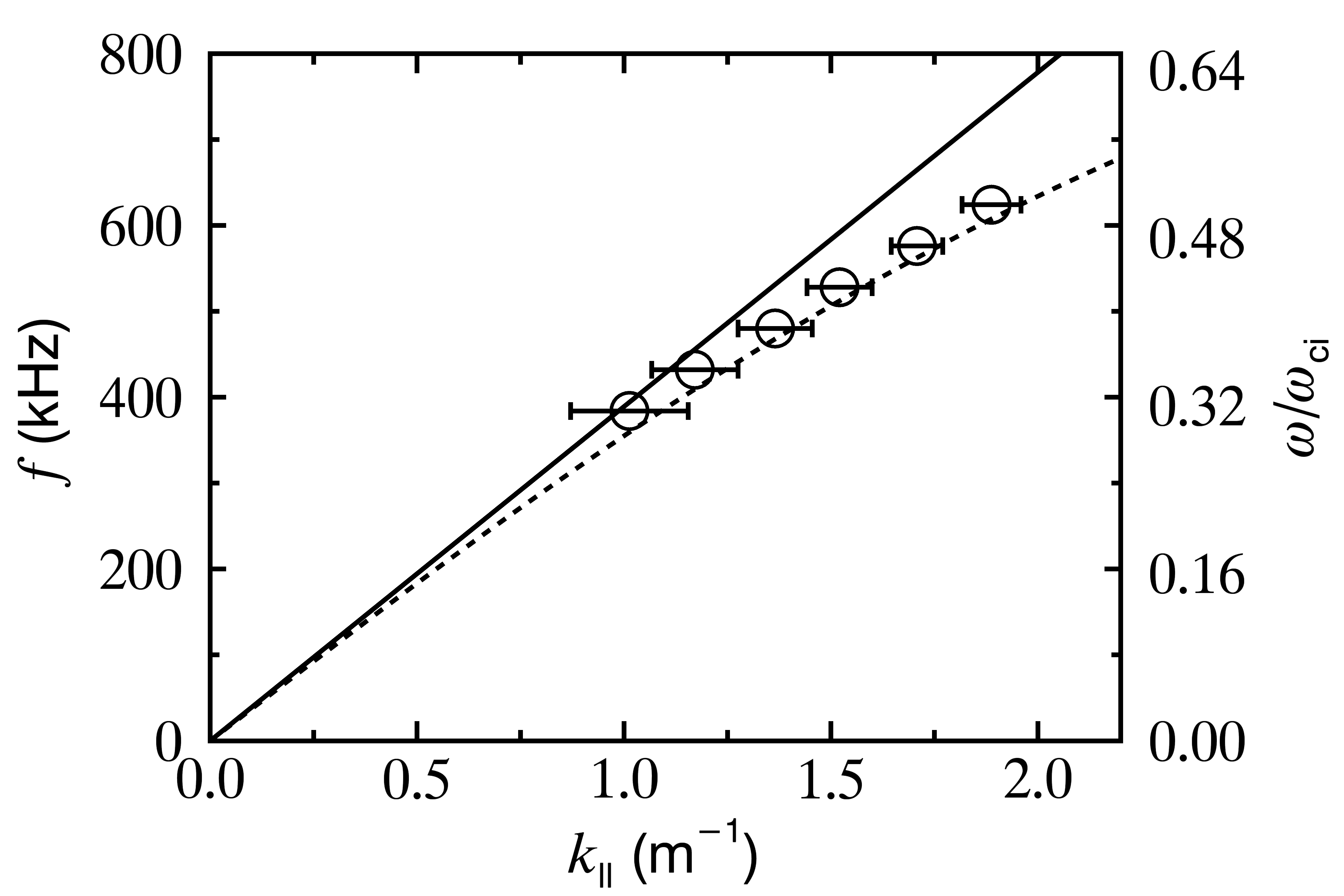}
\caption{Alfv\'en wave dispersion relation for the uniform $B_0 = 800$~G measured before the gradient. The symbols represent the experimental data, while the solid line shows the ideal dispersion relation for Alfv\'en waves and the dashed line the dispersion relation for inertial Alfv\'en waves. 
The ratio of the angular frequency to the ion cyclotron frequency is given on the right-hand axis.
\label{fig:dispersion}}
\end{figure}

Using these results, we can determine the dispersion relation. We show this in Fig.~\ref{fig:dispersion} for the uniform field, $B_0=800$~G case before the gradient, using the $b_y$ component averaged over $x$ from $-6$ to $+6$~cm to avoid any artifacts from the current channels. Also shown in this figure is the ideal dispersion relation for Alfv\'en waves, $\omega/k_{\rm \|}=v_{\rm A}$, and for inertial Alfv\'en waves, which is given by \citep{Stasiewicz:SSR:2000,Gekelman:PoP:2011}
\begin{equation}
\label{iaw_disp}
    v_{{\rm ph},\|} \equiv
    \frac{\omega}{k_{\rm \|}}=v_{\rm A}\frac{\sqrt{(1-\omega^2/\omega_{\rm ci}^2)}}{\sqrt{1+k_{\rm \perp}^2\delta_{\rm e}^2}}. 
    %\frac{\sqrt{}}{}}
\end{equation}%(1+k_\perp^2\rho_i^2)
The perpendicular wave number, \rev{$k_\perp$}, calculated using a Fourier-Bessel analysis \citep{churchill:Book:1987, Vincena:phd:1999}, is $\approx  0.23$~cm$^{-1}$. The data points follows the inertial Alfv\'en wave dispersion relation. In general, the measured $v_{\rm ph,\|}$ for all magnetic field configurations both before and after the gradient fell between $v_{\rm A}$ and the value given by Eq.~(\ref{iaw_disp}). 

\subsection{Wave energy and power}
\subsubsection{Two-dimension measurements}

The total wave energy at a point $(x,y,z)$ and time $t$ can be calculated using the Poynting vector, which is given by \citet{Karavaev:PoP:2011} as 
% \begin{equation} \label{poynting}
%S= \frac{1}{\mu_{\rm 0}}\vec{E} \times \vec{B}
%\end{equation}
\begin{equation} \label{poynting}
S(x,y,z,t)= \frac{1}{\mu_{\rm 0}}b^2v_{\rm g,\|} \approx \frac{1}{\mu_{\rm 0}}b^2v_{\rm ph,\|},
\end{equation}
where $b = \sqrt{b_x^2 + b_y^2 + b_z^2}$ and $v_{\rm g,\|} = \partial \omega/\partial k_{\|}$ is the parallel group velocity. We have made the approximation that $v_{\rm g,\|}$ is approximately equal to the measured $v_{\rm ph,\|}$.    %The measured value falls between $v_{\rm A}$ and $v_{\rm ph,\|}$ from Eq.~(\ref{iaw_disp}), as can be seen in Fig.~\ref{fig:dispersion}. Thus, our approximation introduces maximum errors of $\approx 14\%$ for the lowest frequency and $\approx 24\%$ for the highest frequency. 

The total wave energy in a cross section of LAPD at $z$ can be expressed as
\begin{equation} 
\label{energy}
\mathcal E(z) = \int\int\int S~dx~dy~dt = \frac{v_{\rm ph,\|}}{\mu_{\rm 0}}\int\int\int b^2~dx~dy~dt .
\end{equation}
The wave power $\Gamma$ is related to the total energy in the wave pulse by 
\begin{equation} \label{gama}
\Gamma(z) = \frac{\mathcal E}{t_{\rm dur}},
\end{equation}
where $t_{\rm dur}$ is the duration of the wave pulse. 

\subsubsection{One-dimensional measurements}
\label{trans_1d}

We were able to study five magnetic field configurations within the limited machine time available using one-dimensional (1D) measurements of the wave magnetic field.  
\rev{We used a version of equation~(\ref{energy}) modified for line measurements from $x=-30$~cm to $+30$~cm: 
\begin{equation} 
\label{energy_1d}
\mathcal E(z) = \int\int S~dx~dt = \frac{v_{\rm ph,\|}}{\mu_{\rm 0}}\int\int b^2~dx~dt .
\end{equation} }
We verified the accuracy of this approach by using our two-dimensional (2D) data (e.g., Fig.~\ref{fig:wave_structure}) and extracting an $x$-line measurement at $y=0$. \rev{We then compared the transmittance from $z_1$ to $z_3$ for both the 2D and 1D cases.} The 2D transmission ratio is given by
\begin{equation}
T^{\rm 2D}_{31} =
\frac{\Gamma_3^{\rm2D}}{\Gamma_1^{\rm 2D}}.
\end{equation}
The 1D transmission ratio is \rev{derived from equations~(\ref{gama}) and (\ref{energy_1d})}, giving,
\begin{equation}
T^{\rm 1D}_{31} =
\frac{\Gamma_3^{\rm1D}}{\Gamma_1^{\rm 1D}}.
\end{equation}
We compared the results for a uniform 1600~G case and an $800-2000$~G gradient case and found good agreement between $T^{\rm 2D}_{31}$ and $T^{\rm 1D}_{31}$. We then performed experiments that only measured a line and compared those results to the 2D measurements. Again, we found good agreement for both cases.

%. ADD SOME QUANTITATIVE DISCUSSION TO THE PRECEDING PARAGRAPH. E.G. LINE MEASUREMENT TRANSMITTANCES AGREED WITH PLANE MEASUREMENTS TO WITHIN 5 PERCENT ETC. 
%The transmission for the plane is then compared with the wave transmission from a line measurement and found to be within the uncertainties. We will discuss wave parallel phase speed measurement in a line and compare it with the electron thermal speed, wave transmission in the next subsection \ref{trans}.
\subsection{Transmission}
 
\begin{figure}[ht!]
\centering
\includegraphics[scale=0.38]{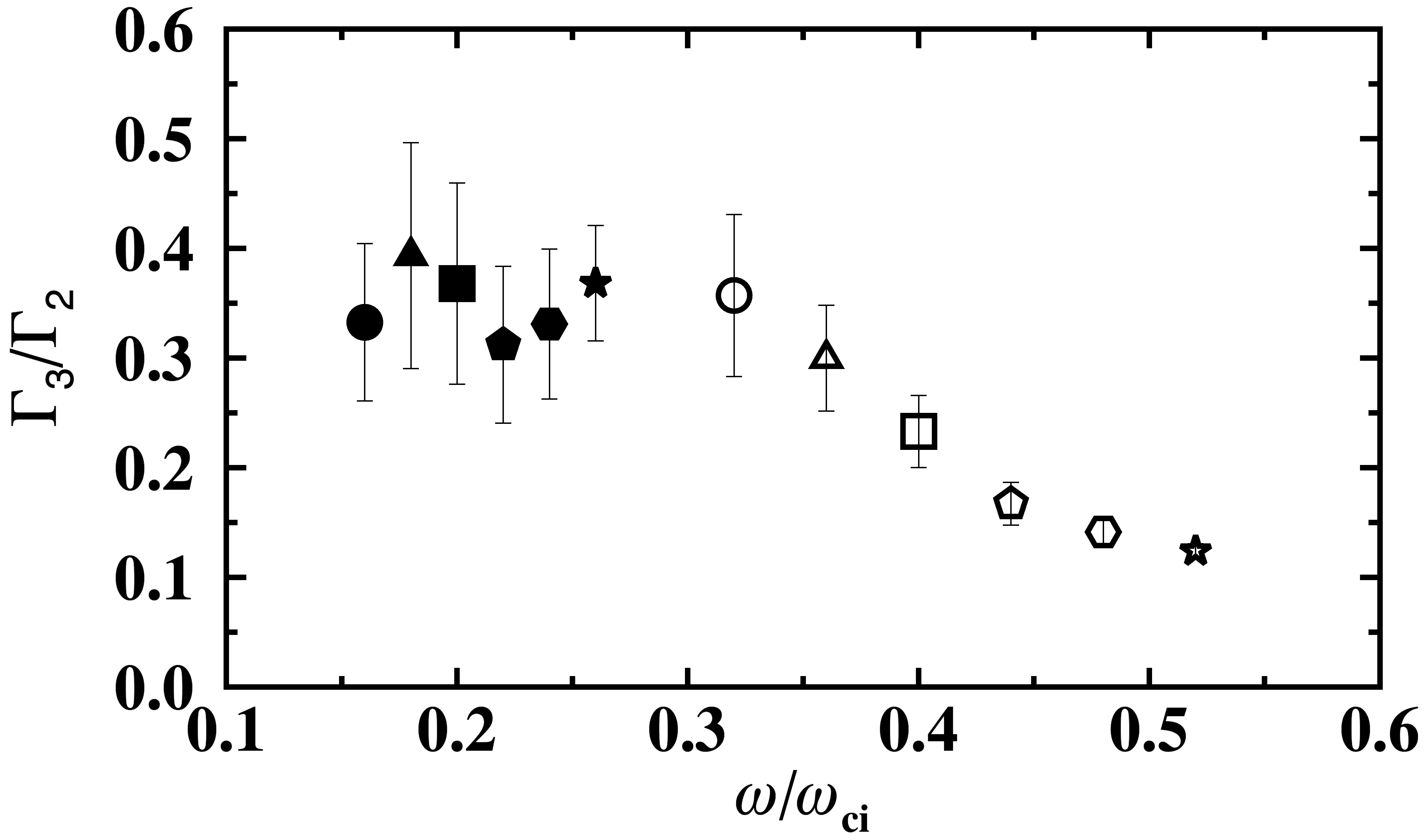}
\caption{1D Measurements of the wave transmission from $z_2$ to $z_3$ for a uniform magnetic field. The open symbols are for $B_0=800$~G and the filled symbols for $1600$~G. The symbol shapes of circle, triangle, square, pentagon, hexagon, and star correspond respectively to frequencies of  390, 439, 488, 536, 585, and 634~kHz. 
\label{fig:damp2}}
\end{figure}

We measured the wave transmission from $z_2$ to $z_3$, $\Gamma_3/\Gamma_2$, for uniform fields of $B_0=800$ and $1600$~G, as shown in Figure~\ref{fig:damp2}. The 1D transmission measurements are plotted as a function of $\omega / \omega_{\rm ci}$. Past LAPD studies have shown that the Alfv\'en waves launched can be reflect from the plasma source \citep{Leneman:PoP:2007}. In order to avoid any potential effects of such reflection, we calculated the travel time of the reflected waves and only used the portion of the incident wave that occurred before the calculated arrival of the reflected wave. The data shown in Fig.~\ref{fig:damp2} were calculated using only the first 2 wave periods for each frequency and field case. This integration time was set by the uniform 1600~G case, for which $v_{\rm A}$ was faster than the 800~G case.

Figure \ref{fig:damp2} shows that the wave energy is significantly damped above $\omega \approx 0.36\omega_{\rm ci}$. We remind the reader that we are using the $\omega_{\rm ci}$ for a pure H$^+$ plasma, for which we would expect to see the onset of ion cyclotron damping above about $0.95\omega_{\rm ci}$ \citep{Vincena:PoP:2001}. \rev{  As the frequency continues to increase, the measured transmission falls further and further below the theoretical prediction , primarily due to the collisional damping \citep{Morales:PoP:1994,Kletzing:PRL:2010}}. This strongly suggests that there are additional ion species in the plasma producing damping. Possible species include H$_2^+$, of which there is likely a significant population in the discharge due to electron impact ionization of the parent H$_2$ gas, and H$_3^+$, which is rapidly formed by the reaction ${\rm H_2^+ + H_2 \to H_3^+ + H}$ \citep{Savic:2020:CPPC}.  The ion cyclotron frequency for H$_2^+$ and H$_3^+$ are, respectively, half and a third of that for H$^+$. In order to study the damping due solely to the gradient, we have limited our analysis to waves with frequencies $\omega \leq 0.36\omega_{\rm ci}$.

\begin{figure}[ht!]
\centering
%\plotone{transmission1.png}
\includegraphics[scale=0.38]{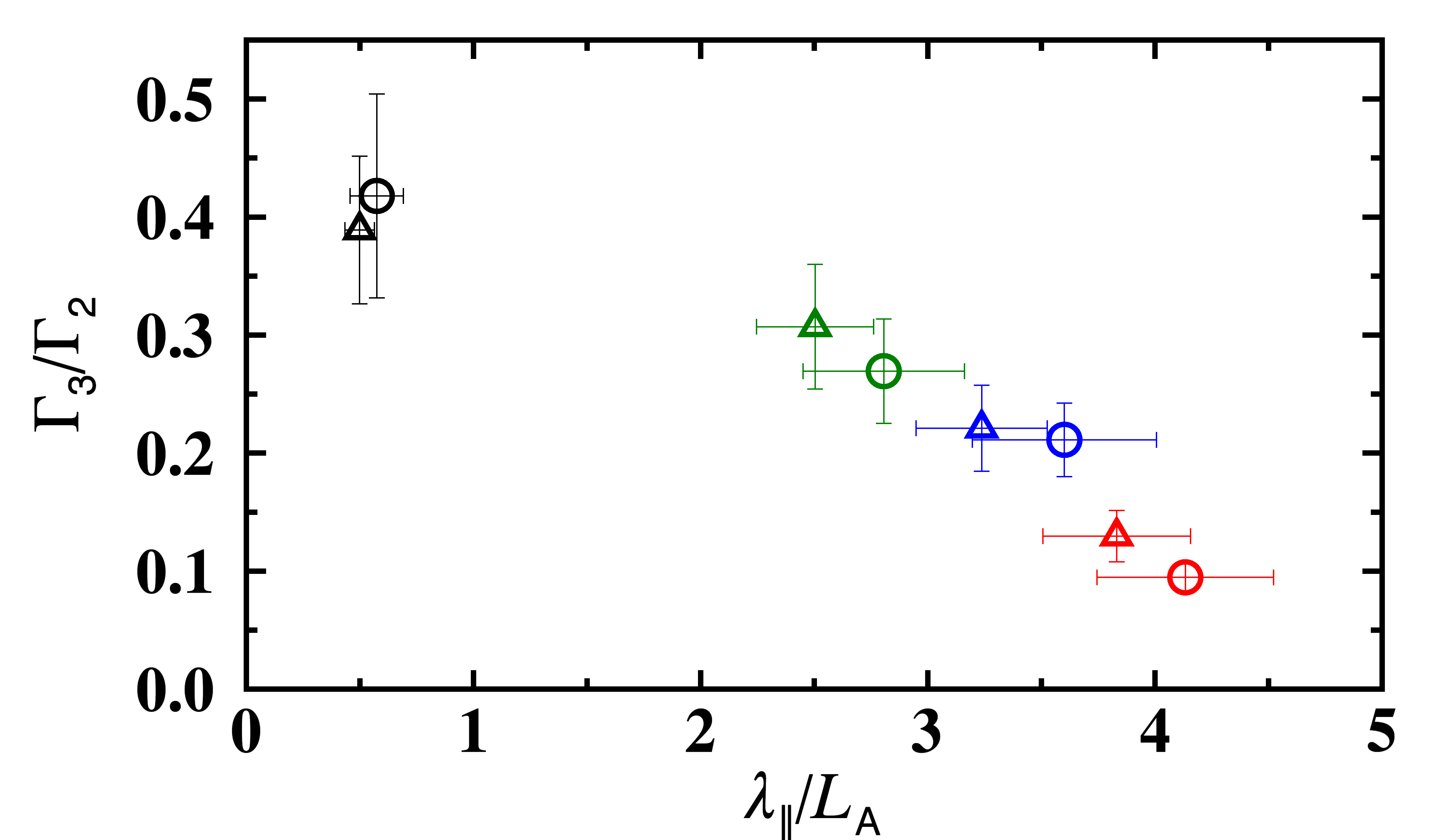}
\caption{Measured 1D wave transmission for a uniform magnetic field $B_0=800$~G (black) and the three gradient cases of  $B_0=800-1600$~G (green),  $B_0=800-2000$~G (blue), and  $B_0=800-2400$~G (red). The symbols are defined in Fig.~\ref{fig:damp2}. 
\label{fig:trang3g2}}
\end{figure}

In Fig.~\ref{fig:trang3g2}, 
we show our measured wave transmission for the uniform $B_0=800$~G case and the three gradient cases. The transmission through the gradient is plotted as a function of $\lambda_\parallel/L_{\rm A}$. The data are shown for each axial magnetic field configuration using the minimum of $L_{\rm A}$ for that profile. Here, we were able to integrate over 5 wave periods, due to the reflected wave traveling slower in these four cases as compared to the uniform 1600~G case.  \rev{Our results show a reduction in the wave energy even in the uniform magnetic field case. We attribute this to collisional damping.} The transmission continues to decrease as the gradient becomes steeper. Note that the difference in the values of the $B_0 =800$~G data in the Figs. \ref{fig:damp2} and \ref{fig:trang3g2} is due to the selection of wave period for the energy calculations.

\begin{figure}[ht!]
\centering
\includegraphics[scale=0.39]{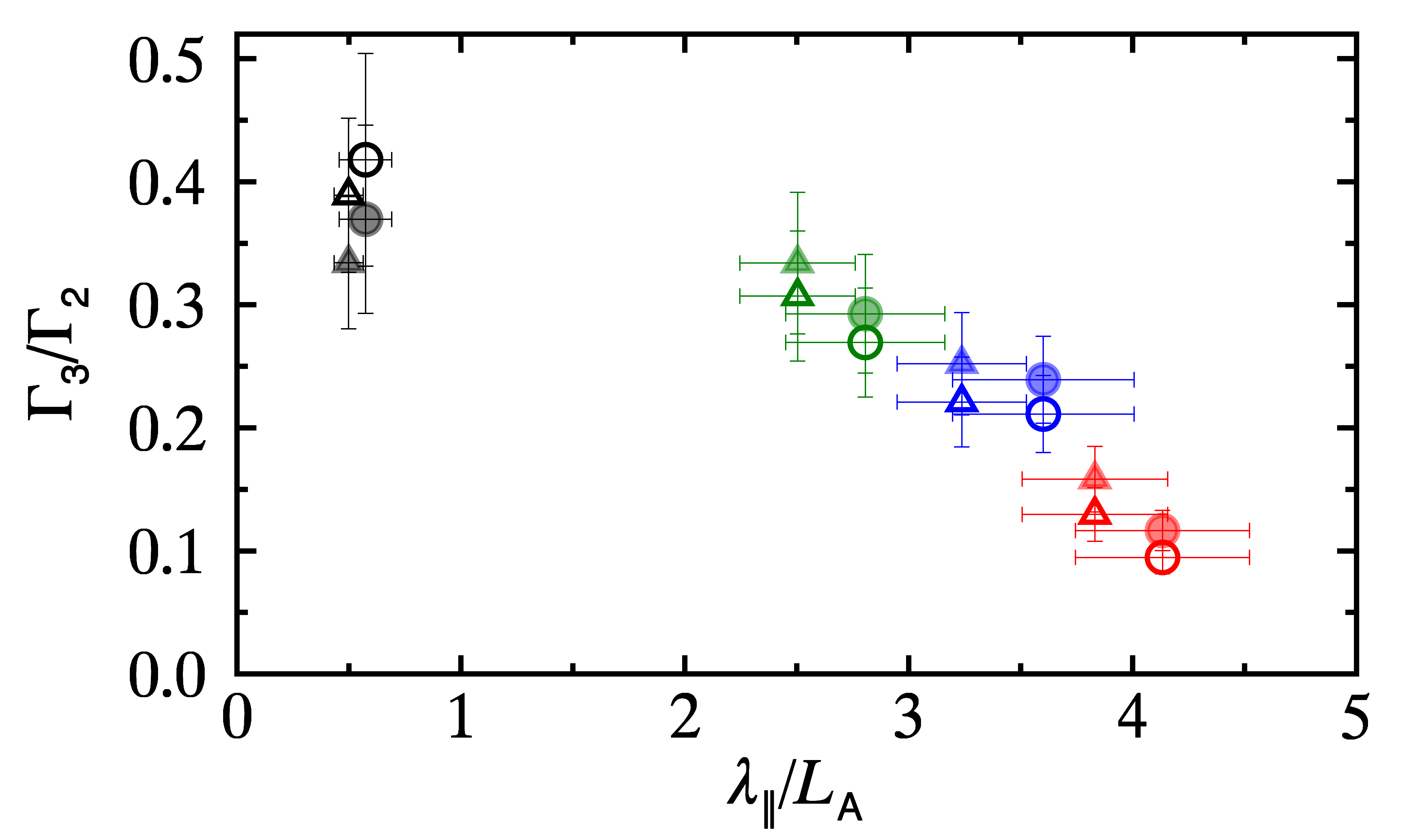}
\caption{\rev{Measured 1D wave transmission over the full range (unfilled symbols) and over the smaller range from  $x= -5$~cm to $+5$~cm (filled symbols). The symbols are defined in Fig. 5 and the colors in Fig. 6..}
\label{fig:trang3g2_5cm}}
\end{figure}

\rev{The effect of phase mixing is ruled out by comparing and calculating wave transmission only within the region where $v_A$ is uniform ($x= -5$~cm to $+5$~cm) to that over the entire measured line. Removing any effect of phase mixing would significantly increase the transmission. Figure \ref{fig:trang3g2_5cm} shows that the calculated transmission in the narrower range increases the transmission of the measurement by an amount ($\sim 10\%$) that is within the error bars of the respective measurements.}
 
\section{Discussion and summary} \label{subsec:dis}
We have performed an experiment to study the propagation of inertial Alfv\'en waves through a longitudinal gradient in the Alfv\'en speed. We launched Alfv\'en waves of six different frequencies from 0.32 to 0.52$\omega_{\rm ci}$. Our gradient transmission measurements are limited to the lowest two frequencies of 0.32 and 0.36$\omega_{\rm ci}$.%, because at higher frequencies we observed strong damping, which we ascribe to ion cyclotron resonance with hydrogen molecular ions. 

 The waves with $\omega \gtrsim 0.36$~$\omega_{\mathrm{ci}}$ have additional damping, which we attribute to ion cyclotron resonance damping. This is likely due to resonance with H$_{2}^+$ and H$_3^+$ ions, whose cyclotron frequencies are, respectively, half and a third that of the H$^{+}$ cyclotron frequency. Ion cyclotron damping also plays a role in coronal heating  \citep{Cranmer:ApJ:2000}. However, in our hydrogen plasma the presence of H$_2^+$ and H$_3^+$ was unintentional and the ion fractions are unknown. Future work could study the characteristics of ion cyclotron damping in a mixed species plasma using known fractions of heavy ions.

\begin{figure}[ht!]
\centering
\includegraphics[scale=0.38]{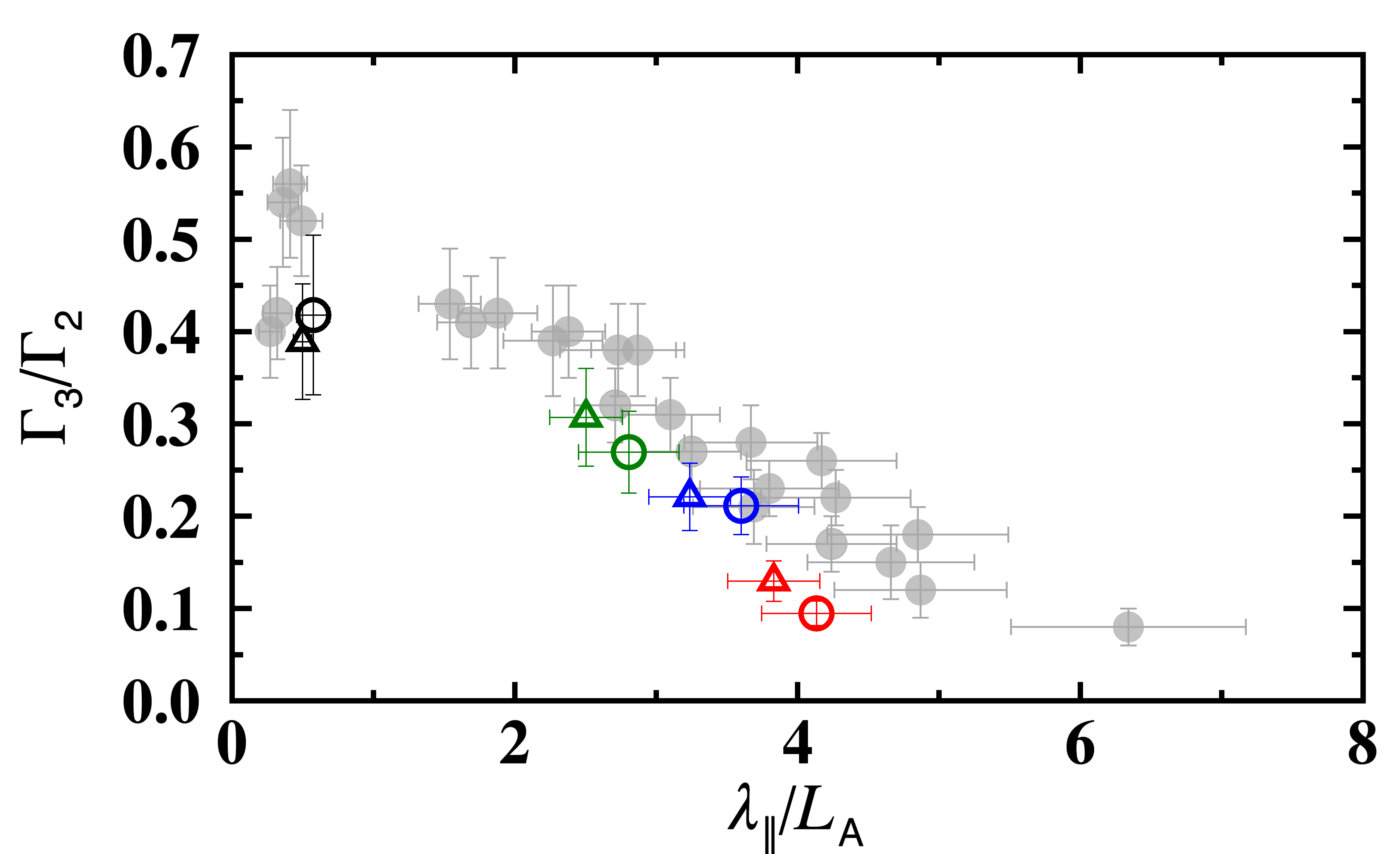}
\caption{ Comparison of our inertial Alfv\'en wave transmission data to the  kinetic Alfv\'en wave results of \citet{Bose:ApJ:2019}, which are shown in grey. The inertial Alfv\'en wave symbols are defined in Figure \ref{fig:trang3g2}.
\label{fig:iaw_kaw}}
\end{figure}

Returning to the lowest frequencies that were unaffected by the resonant damping, we found that the inertial Alfv\'en wave transmission scales with $\lambda_\|/L_{\mathrm{A}}$, just as was found for kinetic Alfv\'en waves \citep{Bose:ApJ:2019}. Furthermore, our experiments demonstrate that Alfv\'en speed gradients reduce the transmission of inertial Alfv\'en wave energy. Figure~\ref{fig:iaw_kaw} compares our present inertial Alfv\'en wave results with the kinetic Alfv\'en wave transmittances measured by \citet{Bose:ApJ:2019}. Comparing these results shows that the transmittances match not only in terms of the scaling, but also in magnitude. That is, the amount of energy lost by both kinetic and inertial Alfv\'en waves when passing through a given Alfv\'en speed gradient is about the same. The transmittance of Alfv\'en waves through a gradients appears to be largely independent of whether the waves are in the kinetic or inertial regime, at least for frequencies well below $\omega_{\mathrm{ci}}$. 

For these inertial regime experiments, we expected Landau damping to be negligible. However, collisional damping continues to play a significant role. The enhanced collisional damping associated with the lower $T_{\mathrm{e}}$ for this experiment likely mitigated the reduced Landau damping, resulting in similar damping to what was observed in the kinetic regime by \citet{Bose:ApJ:2019}. 

These results may find implications in solar physics wherever inertial Alfv\'en waves encounter gradients in the Alfv\'en speed. One example is that Alfv\'en waves generated in solar flares may propagate downwards, reflect off the chromosphere, and thereby produce turbulence by accelerating electrons \citep{Fletcher:aa:2008}. The gradient length scales at the chromosphere imply that $\lambda_\|/L_{\mathrm{A}}$ is larger than in our experimental configurations. Nevertheless, our results show that even at moderate values of the inhomogeneity parameter, the wave transmission is dramatically reduced by the gradient. This suggests that the majority of the coronal Alfv\'en wave energy is not transmitted, but likely undergoes reflection or mode conversion instead. \rev{Coronal Alfv\'en wave energy could be absorbed by the plasma, resulting in heating and/or particle acceleration, in addition to being transmitted, reflected, or mode converted.} 

Any reflected wave energy was below the detection limits for our probes. This implies that the inertial Alfv\'en waves are damped at the gradient. The wave energy may get absorbed near the gradient due to mode conversion \citep{Stix:Book:1992}. We would expect wave damping to eventually deposit energy into the plasma, leading to heating. However, for the small amplitudes used here, the expected change in the electron temperature would be too small to measure \citep{Bose:ApJ:2019}. In future work, we plan to excite larger amplitude waves such that $ b/B_{0} \sim 0.01-0.1$, so that we can quantify the reflected wave energy. Such large amplitude waves are expected to also result in significant plasma heating, measurements of which would help us to better understand the wave dissipation processes in the gradient region. Moreover, larger amplitudes will be more similar to solar Alfv\'en waves. In future studies, we plan to conduct detailed wave and plasma measurements within the gradient region to uncover the physical processes and the unidentified damping mechanism responsible for the wave losing most of its energy.

\section*{Acknowledgment} 
\begin{acknowledgments}
This work was supported by Department of Energy (DOE) grants DE-SC0021261 \rev{and DE-SC0025366}. The experiments were performed at the Basic Plasma Science Facility (BaPSF), which is a collaborative research facility supported by the DOE and the National Science Foundation (NSF). \rev{Garima Joshi was partially supported by DOE under award number DE-SC0022153. Sayak Bose was supported by the National Science Foundation award AGS-2401110. }
\end{acknowledgments}

%\bibliographystyle{aasjournal}
%\bibliography{references}{}

\end{document}